\documentclass[utf8]{FrontiersinHarvard}
\usepackage{url,hyperref,lineno,microtype,subcaption, multirow, array, mathtools, siunitx}
\usepackage{verbatim,amsmath,amsfonts,amssymb,graphicx}
\usepackage[onehalfspacing]{setspace}

\def\vt{v_{T}}
\def\vped{v_{\mathrm{Ped}}}
\def\vtvped{\vt/\vped}
\def\sfwidth{0.8\textwidth}
\def\NEdefval{\infty}
\def\EEdefval{4}

\def\MEdefval{1}
\def\HEdefval{0.4}

\def\NEdefs{u_T \rightarrow \NEdefval}
\def\EEdefs{u_T = \EEdefval}

\def\MEdefs{u_T = \MEdefval}
\def\HEdefs{u_T = \HEdefval}

\def\NEE0{0}
\def\EEE0{24}
\def\LEE0{31}
\def\MEE0{94}
\def\HEE0{235}
\def\VHEE0{941}
\def\mvpm{\unit[per-mode = symbol]{\milli\volt\per\meter}}
\def\NEcase{E_0 = \qty{\NEE0}{\mvpm}}
\def\EEcase{E_0 = \qty{\EEE0}{\mvpm}}

\def\MEcase{E_0 = \qty{\MEE0}{\mvpm}}
\def\HEcase{E_0 = \qty{\HEE0}{\mvpm}}

\def\vthx{v_{th,x}}
\def\vthy{v_{th,y}}

\def\vthj{v_{th,j}}
\def\vybulk{\left\langle v_y \right\rangle}

\def\ftheo{F^{\mathrm{Theory}}}
\def\fptheo{F^{\prime\mathrm{Theory}}}

\def\ftheoy{\ftheo_y}
\def\ftheoz{\ftheo_z}

\def\fptheoy{\fptheo_y}

\def\leftlim{\qty[per-mode = symbol]{-20}{\kilo\meter\per\second}}
\def\rightlim{\qty[per-mode = symbol]{20}{\kilo\meter\per\second}}
\def\mps{\unit[per-mode = symbol]{\meter\per\second}}
\def\dv{\qty[per-mode = symbol]{78.125}{\mps}}

\def\keyFont{\fontsize{8}{11}\helveticabold }
\def\firstAuthorLast{Koontaweepunya {et~al.}}
\def\Authors{Rattanakorn Koontaweepunya\,$^{1,*}$, Yakov S. Dimant\,$^{1}$ and Meers M. Oppenheim\,$^{1}$}
\def\Address{$^{1}$Center for Space Physics, Boston University, Boston, MA, United States}

\def\corrAuthor{Rattanakorn (Save) Koontaweepunya}

\def\corrEmail{rakoon@bu.edu}

\begin{document}
\onecolumn
\firstpage{1}

\title[Non-Maxwellian Electrojet Ions]{Non-Maxwellian Ion Distribution in the Equatorial and Auroral Electrojets}

\author[\firstAuthorLast ]{\Authors} 
\address{} 
\correspondance{} 

\extraAuth{}

\maketitle

\begin{abstract}

\section{}

Strong electric fields in the auroral and equatorial electrojets can distort the background ion distribution function away from Maxwellian. We developed a collisional plasma kinetic model using the Boltzmann equation and a simple BGK collision operator which predicts a relatively simple relationship between the intensity of the background electric field and the resulting ion distribution function. To test the model, we perform 3-D plasma particle-in-cell simulations and compare the results to the model. The simulation applies an elastic collision operator assuming a constant ion-neutral collision rate. These simulations show less ion heating in the Pedersen direction than the analytic model but show similar overall heating. The model overestimates the heating in the Pedersen direction because the simple BGK operator includes no angular collisional scattering in the ion velocity space. On the other hand, the fully-kinetic particle-in-cell code is able to capture the physics of ion scattering in 3-D and therefore heats ions more isotropically. Although the simple BGK analytic theory does not precisely model the non-Maxwellian ion distribution function, it does capture the overall momentum and energy flows and therefore can provide the basis of further analysis of E-region wave evolution.

\tiny
 \keyFont{ \section{Keywords:} E-region electrojet, ion distribution function, BGK collision operator, Maxwell molecule collision model, Pedersen conductivity, PIC simulation, plasma instabilities, ion temperature anisotropy} 
\end{abstract}

\section{Introduction}

Strong DC electric fields in the auroral and equatorial electrojets drive plasma instabilities in the E-region ionosphere. When perpendicular to the global magnetic field, these electric fields generate strong cross-field plasma instabilities, such as the Farley-Buneman instability \citep{Farley1963, Buneman1963}, the gradient drift instability \citep{Hoh:63, Maeda1963, Simon:Instability63}, the electron thermal instability \citep{Dimant_Sudan:1995b, Dimant_Sudan:1997, Robinson:effects98, StMaurice:role00}, and the ion thermal instability \citep{Kagan:thermal00, Dimant_Oppenheim:2004}. These plasma instabilities serve to explain the plasma density irregularities that for many years have been observed in the E-region ionosphere by radars and sounding rockets.

Analytic models of plasma instabilities often employ the kinetic description which accurately describes plasma as individual particles. Despite the accuracy of these kinetic models, they are too complicated and often impractical to work with. The complexity of collisional kinetic plasma models make it challenging to simulate plasma on a large scale. Kinetic simulations of plasma using Particle-in-Cell (PIC) codes can take very long time to simulate very little. For example, \cite{Oppenheim:Large08} spent over four years of CPU time to simulate a 2-D patch of plasma for a quarter of a second. Other PIC simulations by \cite{Oppenheim_Dimant:2004}, \cite{Oppenheim_Dimant:2013}, and \cite{Oppenheim2020} also exhausted large amounts of computer resources on many different supercomputers. It is therefore practical to develop a fluid model which is computationally efficient while at the same time able to capture the kinetics of plasma that affects the development of plasma instabilities.

To develop an accurate fluid model applicable for accurate description of plasma instabilities in the E-region ionosphere, we need to understand how the electric fields in the electrojets affect the background ion distribution function. The E-region ionosphere is weakly ionized, with neutrals outnumbering ions $10^6$ to one \citep{Schunk:Ionospheres09}. In the lower E-region, the ions are essentially unmagnetized by frequent collisions with neutrals \citep{Kelley:Ionosphere09}. These collisions prevent ions from accelerating ad infinitum along the electric field. As a result, in steady state, the bulk of the ions drifts on average with the Pedersen velocity, which is proportional to the electric field divided by the ion-neutral collision rate.

Due to the ion Pedersen drift, the background ions in the E-region ionosphere may not perfectly follow the Maxwellian distribution characteristic for the plasma in thermal equilibrium \citep{Chen:Plasma16}. If the external DC electric field is strong enough, it leads to an anisotropy in the ion distribution function with clear distortions from Maxwellian. \cite{StMaurice1979} gives the theory and experimental evidence for non-Maxwellian ion distribution functions in the auroral E- and F-regions. The DC electric field can be especially strong at high latitudes during intense geomagnetic effects (storms). Compared to the high-latitude E- and F-regions, the equatorial E-region has less intense electric fields, so we expect the distortion in the ion distribution to be small. However, extreme geomagnetic storms can intensify the electric fields even there, so that there may be periods where the ion distribution function deviates significantly from Maxwellian even in the equatorial E-region.

In this paper, we develop a collisional plasma kinetic model which relates the intensity of the external electric field to the ion velocity distribution function. We restrict our treatment to a spatially uniform and quasi-steady ionosphere, which represents the background for developing instabilities. To describe the ion-neutral collisions, our kinetic model uses the BGK collision operator \citep{Bhatnagar:Model54}, which is a crude way of describing plasma collisions \citep{Nicholson:Plasma83}. Despite its inaccuracy, however, this simplified operator conserves the particle number, the total momentum and energy of the colliding particles. A hybrid simulation by \cite{Kovalev:Modeling09}, based on the BGK collision term for ions, was able to give results comparable to the more accurate hybrid and full PIC simulations \citep{Janhunen1995, Oppenheim1995, Oppenheim1996a, Oppenheim:Large08, Oppenheim_Dimant:2004} despite. \cite{Else2009} found that the constant collision rate BGK model is in agreement with a more realistic constant mean free path model in regimes where the Pedersen velocity is less than or comparable to the neutral thermal velocity. In this paper, we compare the analytic results based on the oversimplified BGK model to the more accurate model of a fully-kinetic PIC code. 

The paper is organized as follows. Section \ref{sec:methods} describes the simulation methods. Section \ref{sec:results} presents the analytic model and compares it to the simulation results. Section \ref{sec:disscussion} discusses the discrepancies between the analytical results and the simulation. Section \ref{sec:conclusions} concludes the paper.

\section{Simulation Methods}
\label{sec:methods}

We used EPPIC --- the Electrostatic Parallel Plasma-in-Cell Simulator --- to simulate the E-region background ions. EPPIC, like other particle-in-cell (PIC) codes, simulates plasma as individual particles. The existence of particles enables PIC simulations to reproduce the kinetic behaviors of plasma. We are interested in the kinetic behavior of plasma that is the distortion of the ion distribution function. For more information about PIC codes, see \cite{Birdsall:Plasma}. For detailed explanations of EPPIC, see \cite{Oppenheim_Dimant:2004}, \cite{Oppenheim:Large08}, and \cite{Oppenheim_Dimant:2013}.

We set the magnetic field to zero in our simulation, because the E-region background ions are unmagnetized. We also exclude electrons from our simulation, using instead a uniform background electron plasma that does not respond to any fields. We do this because this paper only explores the physics of the ion distribution function, independent of the electron generated fields. EPPIC simulates background ions as PIC particles and neutrals as a uniform, constant background. Our simulation is in three-dimension (3-D), even though a two-dimensional (2-D) simulation would have sufficed for the behavior we were interested in. Table \ref{tab:1} gives the simulation parameters.

The E-region background ions are highly collisional with the neutrals. We use the Maxwell molecule collision model in our simulation. In Maxwell molecule collisions, the ion-neutral collision rate is constant and does not depend on the particle's velocity. \cite{Schunk:Ionospheres09} gives a detailed explanation of the Maxwell molecule collision model. EPPIC employs a statistical method of applying collisional effects to ions. At each time step, EPPIC designates a number of ions to be collided in accordance with the ion-neutral collision rate specified in the input deck. Collision changes the ion's momentum and scatters it elastically in 3-D. Collisions do not affect neutrals in our simulation, since the simulation treats the neutrals as a uniform, constant background. In the E-region, neutrals are many orders of magnitude more numerous than ions $\left(n_n/n_i > 10^6; \text{ \cite{Schunk:Ionospheres09}} \right)$; therefore, neutrals that collide with ions constitute a very small part of the neutrals and do not affect the overall momentum and temperature of the neutrals.

Section \ref{sec:PIC} details the specific simulation setup as well as the as the analysis methods used for the simulation results.

\begin{table}
\caption{Simulation parameters}
\centering
\setlength{\extrarowheight}{3pt}
\begin{tabular}{|l|c|c|}
\hline 
Simulation Parameter & Symbol & Value \\ 
\hline 
\multicolumn{3}{|l|}{\textit{Ion Parameters}} \\ 
\hline 
Ion mass 							& $m_i$ 								& \qty{5e-26}{kg} \\ 
\hline 
Ion-neutral collision rate 	& $\nu_{in}$ 						& \qty{1050}{s^{-1}} \\ 
\hline 
Ion number density 			& $n_0$ 								& \qty{4e8}{m^{-3}} \\ 
\hline 
Ion charge						& $e$ 									& \qty{1.602e-19}{C}\\ 
\hline
\multicolumn{3}{|l|}{\textit{Neutral Parameters}} \\
\hline 
Neutral thermal velocity	& $\vt$ 								& \qty{287}{\mps} \\ 
\hline 
Neutral mass 					& $m_n$ 								& \qty{5e-26}{kg} \\ 
\hline 
\multicolumn{3}{|l|}{\textit{Simulation Parameters}} \\
\hline
Grid size							& $dx = dy = dz$					& \qty{0.15}{m} \\
\hline
Number of grids				&	$\left( nx, ny, nz \right)$		& (1024, 512, 512) \\	
\hline
Time step 						& $dt$ 									& \qty{5.6e-5}{s} \\ 
\hline 
Number of time steps 		& $nt$ 									& 512 \\ 
\hline 
\end{tabular}
\label{tab:1}
\end{table}

\section{Results}
\label{sec:results}

\subsection{Analytic Model of the Background Ion Distribution Function} \label{sec:yakov}

\subsubsection{Derivation of the Distorted Ion Distribution Function}

The simplest kinetic equation for the ion distribution function (IDF) with the
Bhatnagar-Gross-Krook (BGK) collision term \citep{Bhatnagar:Model54} is given by
\begin{equation}
\frac{\partial f}{\partial t}+\frac{e}{m_{i}}\ \vec{E}\cdot\frac{\partial
f}{\partial\vec{V}}+\vec{v}\mathbf{\,\cdot\,}\frac{\partial f}{\partial\vec
{r}}=-\;\nu_{in}\left(  f-\frac{n_{i}(\vec{r},t)}{n_{0}}\ \ f_{0}%
^{\mathrm{Coll}}\right)  , \label{initial_ion_kinetic_equation}%
\end{equation}
where $v=|\vec{v}|$ is the ion speed, $\nu_{in}$ is the ion-neutral collision
frequency, $T_{n}$ is the neutral temperature (in energy units), $m_{i}$ is
the ion mass (equal to the neutral mass), $\vec{E}$ is the external electric field, and
\begin{equation}
f_{0}^{\mathrm{Coll}}\left(  v\right)  \equiv n_{0}\left(  \frac{m_{i}}{2\pi
T_{n}}\right)  ^{3/2}\exp\left(  -\ \frac{m_{i}v^{2}}{2T_{n}}\right)  .
\label{ion_cooling _Maxwellian}%
\end{equation}
The function $f_{0}^{\mathrm{Coll}}\left(  v\right)  $ is the spatially
uniform and stationary ion Maxwellian distribution function, normalized to the
mean ion density $n_{0}$ with no external electric field. The
BGK collision term on the RHS of Eq.~(\ref{initial_ion_kinetic_equation})
assumes Maxwell collisions \citep{Schunk:Ionospheres09} with the given constant $\nu_{in}$.

Below we consider only the background conditions with the externally imposed
electric before developing any instabilities, $\vec{E}=\vec{E}_{0}$. For the
corresponding spatially uniform and stationary background ion distribution
function $f_{0}(\vec{v})$, Eq.~(\ref{initial_ion_kinetic_equation}) reduces to%
\begin{equation}
\vec{a}_{0}\cdot\frac{\partial f_{0}}{\partial\vec{V}}=-\;\nu_{in}\left(
f_{0}-f_{0}^{\mathrm{Coll}}\right)  , \label{bkg_ion_kinetic_equation}%
\end{equation}
where $\vec{a}_{0}\equiv e\vec{E}_{0}/m_{i}$ is the free-ion acceleration.
Introducing a Cartesian coordinate system with the axis $y$ directed along
$\vec{a}_{0}$ and integrating Eq. (\ref{bkg_ion_kinetic_equation}) over the perpendicular velocity components
$v_{y}$ and $v_{z}$, we obtain%
\begin{equation}
a_{0}\ \frac{\partial F_{0}}{\partial V_{y}}=-\nu_{in}\left(  F_{0}%
-F_{0}^{\mathrm{Coll}}\right)  . \label{via_a_0}%
\end{equation}
Here%
\begin{equation}
F_{0}(v_{y})\equiv%
{\displaystyle\iint\limits_{-\infty}^{+\infty}}
f_{0}\,dv_{x}dv_{z} \label{F_0}%
\end{equation}
and
\begin{equation}
F_{0}^{\mathrm{Coll}}(v_{y})\equiv\frac{n_{0}}{\sqrt{2\pi}v_{Ti}}\exp\left(
-\ \frac{v_{y}^{2}}{2v_{Ti}^{2}}\right)  , \label{relaxed_F0}%
\end{equation}
where $v_{Ti}\equiv\sqrt{T_{n}/m_{i}}$ is the thermal velocity of the neutral
particles ($m_{i}=m_{n}$). In the BGK appproximation, the ion velocity
distribution in the two perpendicular directions remains undisturbed by the
field $\vec{E}_{0}$, so that the full 3-D IDF becomes%
\begin{equation}
f_{0}(v_{x},v_{y},v_{z})=\frac{F_{0}(v_{y})}{2\pi v_{Ti}^{2}}\exp\left(
-\ \frac{v_{x}^{2}+v_{z}^{2}}{2v_{Ti}^{2}}\right)  . \label{f_0}%
\end{equation}

Plugging Eq.~(\ref{relaxed_F0}) into Eq.~(\ref{via_a_0}) and solving the
latter yields%
\begin{equation}
F_{0}\left(  v_{y}\right)  =\frac{n_{0}\nu_{in}}{2a_{0}}\exp\left[
-\ \frac{\nu_{in}v_{y}}{a_{0}}+\frac{1}{2}\left(  \frac{\nu_{in}v_{Ti}}{a_{0}%
}\right)  ^{2}\right]  \left[  1+\operatorname{erf}\left(  \frac{v_{y}%
-\nu_{in}v_{T}^{2}/a_{0}}{\sqrt{2}v_{Ti}}\right)  \right]  , \label{F_0(V_x)}%
\end{equation}
where $\operatorname{erf}(y)=(2/\sqrt{\pi})\int_{0}^{y}e^{-t^{2}}dt$ is the
error function. Introducing the dimensionless ion velocity $u\equiv\nu
_{in}v_{y}/a_{0}$ and the dimensionless neutral thermal velocity $u_{T}%
\equiv\nu_{in}v_{Ti}/a_{0}$, we can recast Eq.~(\ref{F_0(V_x)}) as%
\begin{align}
G_{0}(u)  &  =\frac{1}{2}\exp\left(  -u+\frac{u_{T}^{2}}{2}\right)  \left[
1+\operatorname{erf}\left(  \frac{u-u_{T}^{2}}{\sqrt{2}u_{T}}\right)  \right]
\nonumber\\
&  =\frac{1}{2}\exp\left(  -\ \frac{u^{2}}{2u_{T}^{2}}\right)  w\left(
-i\ \frac{u-u_{T}^{2}}{\sqrt{2}u_{T}}\right)  , \label{G_0_via_erf}%
\end{align}
where $G_{0}(u)=[a_{0}/(n_{0}\nu_{in})]F_{0}(v_{y})$ and $w(\zeta
)=e^{-\zeta^{2}}\left[  1+\operatorname{erf}(i\zeta)\right]  $. The function
$w(\zeta)$ can be written in terms of the standard plasma dispersion function,
$Z(\zeta)$, as $w(\zeta)=-(i/\sqrt{\pi})Z(\zeta)$.

The solution in the form of Eq.~(\ref{G_0_via_erf}) automatically conserves
the number of particles and provides the correct expressions for the Pedersen
velocity and effective temperature (see below), as can be deduced from the
following integral relationships:
\begin{equation}
\int_{-\infty}^{+\infty}G_{0}(u)\,du=1,\qquad\int_{-\infty}^{+\infty}%
uG_{0}(u)\,du=1,\qquad\int_{-\infty}^{+\infty}u^{2}G_{0}(u)\,du=u_{T}^{2}+2.
\label{three_lowest_moments}%
\end{equation}

\begin{figure}[ptb]
\begin{center}
\includegraphics[width=0.8\linewidth]{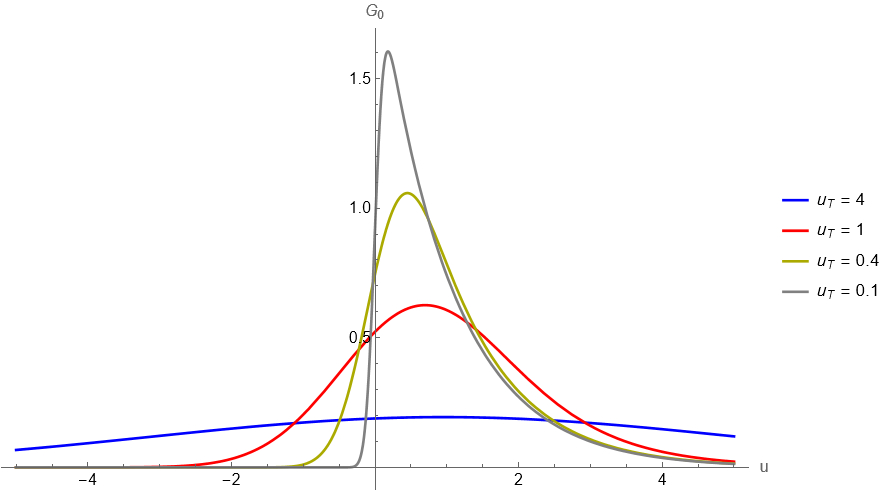}
\end{center}
\caption{The normalized ion distribution function (IDF) for four values of
$u_{T} \equiv\vtvped$, where $ \vped = a_{0}/\nu_{in}$
is the ion Pedersen velocity proportional to $E_{0}$.
The vertical axis is the function $G_{0}(u)$ as seen in Eq.
\eqref{G_0_via_erf}. The horizontal axis is the normalized ion velocity
$u \equiv v_{y}/\vped$. Since $u$ is normalized to
$a_{0}^{-1}$, the IDF is compressed in the horizontal axis by a factor
$\propto E_{0}$; therefore, the effective heating does not relate to the full
width at half maximum (FWHM) in the usual way. In this plot, curves with
smaller FWHM are more strongly heated.}%
\label{fig:1}%
\end{figure}

Figure \ref{fig:1} shows the normalized ion distribution function in Eq.
\eqref{G_0_via_erf} for four values of $u_{T}$. Note that $u_{T} \propto
E_{0}^{-1}$, so the four values of $u_{T}$ in Figure \ref{fig:1} correspond to
four values of $E_{0}$. The ion distribution functions with large values of
$u_{T}$ assume Maxwellian shapes, while the ion distribution functions with
small values of $u_{T}$ appear right-skewed when compared to Maxwellian. The
distortion is such that their peaks lie to the left of their bulk velocity,
which is equal to one according to Eq. \eqref{three_lowest_moments}. Section
\ref{sec:distortionlimits} explains why the ion distribution function retains
the Maxwellian shape at lower values of $u_{T}$ but gets distorted at higher
values of $u_{T}$.

\subsubsection{Distortion of the Ion Distribution Function in the Low and High
$E_{0}$ Limits}

\label{sec:distortionlimits}

The antisymmetric error function, $\operatorname{erf}\left(  \xi\right)  $, at
large $|\xi|$ can be approximated as
\begin{equation}
\operatorname{erf}\left(  \xi\right)  \approx\left\{
\begin{array}
[c]{cc}%
1-\frac{\exp\left(  -\xi^{2}\right)  }{\xi\sqrt{\pi}} & \text{if }\xi>0\text{
and }\xi\gg1\\
-1+\frac{\exp\left(  -\xi^{2}\right)  }{\left(  -\xi\right)  \sqrt{\pi}} &
\text{if }\xi<0\text{ and }\left(  -\xi\right)  \gg1
\end{array}
\right.  .\label{erf(x)_asymptotics}%
\end{equation}
Using the bottom approximation from Eq. \eqref{erf(x)_asymptotics}, we can
show that in the limit where $a_{0}\rightarrow0$,
\begin{equation}
G_{0}(u)\rightarrow\frac{1}{\sqrt{2\pi}u_{T}}\ \exp\left(  -\ \frac{u^{2}%
}{2u_{T}^{2}}\right)  .\label{G_0->}%
\end{equation}
This corresponds to $f_{0}\rightarrow\ f_{0}^{\mathrm{Coll}}$; i.e., the
background ion distribution tends towards Maxwellian in the low $E_{0}$ limit.
Equation \eqref{G_0->} does not hold for all values of $u$. As seen from Eq.
\eqref{G_0_via_erf}, Eq. \eqref{G_0->} does not hold if $u\gg u_{T}^{2}$. This
means that the positive tail of the ion distribution function may deviate
significantly from Maxwellian.

The low $E_{0}$ limit can be expressed in terms of the ion Pedersen velocity,
$\displaystyle \vped = \left\langle v_{y}\right\rangle = a_{0}/\nu_{in} = eE_{0}/(m_{i}\nu_{in})$, and the neutral thermal velocity
$\vt$. If $\vped \ll\vt$, then the distortion to the ion distribution function
is weak, since the ion distribution function tends towards Maxwellian. The
effective temperature,
\begin{equation}
T_{\mathrm{eff}}=T_{n}+\frac{m\vped^{2}}{2}\label{T_eff}%
\end{equation}
is only slightly higher than $T_{n}$, since $m\vped^{2}\ll T_{n}$ in this limit.

In the high $E_{0}$ limit where $\vped\gg\vt$, Eq. \eqref{G_0_via_erf} does
not tend towards Maxwellian, so the ion distribution function will be
distorted along the $\vec{E}_{0}$ direction. The corresponding heating will be
huge as well, since $m\vped^{2}\gg T_{n}$ in Eq. \eqref{T_eff}. Note that the
effective thermal velocity, $\displaystyle\sqrt{T_{\mathrm{eff}}/m_{i}}$, is of the order of the Pedersen velocity: $\displaystyle\sqrt{T_{\mathrm{eff}}/m_{i}}\approx\vped/\sqrt{2}$. 

\subsection{Background Ion Distribution Functions from the PIC Simulation}
\label{sec:PIC}

\subsubsection{Kinetic Simulation of Highly Collisional, Unmagnetized, $\vec{E_0}$-Driven Background Ions}

Our model from section \ref{sec:yakov} predicts the background ion distribution function (IDF) to distort away from Maxwellian when $E_0$ is high enough. Equation \eqref{F_0(V_x)} gives the one-dimensional IDF we expect to see in the $\vec{E_0}$ direction. To test the validity of our model, we run four simulation cases using EPPIC. The values of $E_0$ used in the simulation cases are:
\begin{enumerate}
	\item $\NEcase$, which corresponds to $\NEdefs$.
	\item $\EEcase$, which corresponds to $\EEdefs$.
	\item $\MEcase$, which corresponds to $\MEdefs$.
	\item $\HEcase$, which corresponds to $\HEdefs$.
\end{enumerate}
Like before, $\displaystyle u_T \equiv \vtvped$ is the normalized neutral thermal velocity, $\displaystyle \vt = \sqrt{T_n/m_i}$ is the neutral thermal velocity, and $\displaystyle \vped = e E_0/m_i \nu_{in}$ is the ion Pedersen velocity. 

Our simulation includes one ion species, one neutral species, and no electrons. The imposed electric field $\vec{E_0}$ points in the y direction, and there is no imposed magnetic field. As discussed in Section \ref{sec:methods}, the setup is representative of the plasma condition in the E-region ionosphere where ions are unmagnetized and highly collisional with the neutrals. 

Table \ref{tab:1} gives the parameters used across all simulation cases.

\subsubsection{Normalization of the Discrete Ion Velocity Distribution from the Simulation}
The simulation outputs a $(v_x \times v_y \times v_z) = (512 \times 512 \times 512)$ array of ion velocity distribution. Each dimension of the array covers a velocity domain of [\leftlim, \rightlim]. The grid size is $\displaystyle \Delta v = [\rightlim - (\leftlim)]/512 = \dv$ in each dimension. We reduce the three-dimensional velocity distribution array $f(v_x, v_y, v_z)$ into three one-dimensional velocity distribution arrays: $F_x(v_x)$, $F_y(v_y)$, and $F_z(v_z)$ by summing over two other dimensions. This gives us
\begin{equation}
F_x(v_x) =\sum_{v_y} \sum_{v_z} f(v_x, v_y, v_z),
\end{equation}
and similarly for $F_y(v_y)$ and $F_z(v_z)$. 

To facilitate the comparison with the theory, we normalize $F_x(v_x)$, $F_y(v_y)$, and $F_z(v_z)$ such that the sum of each distribution is equal to $\left(\Delta v\right)^{-1}$. This process is analogous to letting the zeroth velocity moment of a continuous distribution function equal to one. This in effect normalizes the ion number density to one. The normalized arrays are given by
\begin{equation}
F_k'(v_k) = \frac{F_k(v_k)}{\sum_{v_k} F_k(v_k) \Delta v},
\label{fp}
\end{equation}
where $k$ is either $x$, $y$, or $z$. The normalization makes it so that $\sum_{v_k} F_x'(v_x) = \left(\Delta v\right)^{-1}$ for all $k$.

\subsubsection{Normalization of the Continuous Ion Velocity Distribution from the Theory}

The continuous one-dimensional ion distribution function in the direction parallel to $\vec{E_0}$ direction is given by the theory as $F_0(v_y)$ in Eq. \eqref{F_0(V_x)}. For clarity, we reiterate Eq. \eqref{F_0(V_x)} as
\begin{equation}
\ftheoy(v_y)= 
\begin{dcases}
    \frac{n_{0}\nu_{in}}{2a_{0}}
\exp\left[-\ \frac{\nu_{in}v_{y}}{a_{0}}+\frac{1}{2}\left(  \frac{\nu_{in}\vt}{a_{0}}\right)  ^{2}\right]  
\left[  1+\operatorname{erf}\left(  \frac{v_{y}-\frac{\vt^{2}\nu_{in}}{a_{0}}}{\sqrt{2}\vt}\right)  \right], & \text{if } a_0 \neq 0\\
    \frac{n_0}{\sqrt{2 \pi}\vt}\exp\left(-\frac{v_y^2}{2\vt^2}\right),              & \text{if } a_0 = 0
\end{dcases},
\label{ftheoy}
\end{equation}
where we have incorporated the result in the low $E_0$ limit from Section \ref{sec:distortionlimits}.

For the directions perpendicular to $\vec{E_0}$, the theory assumes an undisturbed Maxwellian given by
\begin{equation}
\ftheo_j(v_j) =  \frac{n_0}{\sqrt{2 \pi}\vt}\exp\left(-\frac{v_j^2}{2 \vt^2}\right),
\end{equation}
where $j$ is either $x$ or $z$.

To facilitate the comparison with the simulation results, we normalize $\ftheo_x(v_x)$, $\ftheoy(v_y)$, and $\ftheoz(v_z)$ such that the area under the curve of each distribution is equal to one. This sets the zeroth velocity moment of the distribution to one and normalizes the ion number density to one. The normalized distribution functions are given by
\begin{equation}
\fptheo_k(v_k) = \frac{\ftheo_k(v_k)}{\int_{-\infty}^{\infty}\ftheo_k(v)\,dv},
\label{fptheo}
\end{equation}
where $k$ is either $x$, $y$, or $z$. The normalization makes it so that $ \int_{-\infty}^{\infty} \fptheo_k(v_k) \, dv_k = 1$ for all $k$.

\subsubsection{Choice of $\nu_{in}$ in the Theoretical Results}
\label{sec:nu}

Although EPPIC used the ion-neutral collision rate $\nu_{in} = \qty{1050}{s^{-1}}$ as its input, the outputted $F'_y(v_y)$ exhibits $\nu_{in} = \qty{1082}{s^{-1}}$. The simulation gives the ion bulk velocity $\vybulk$, and the relation $\vybulk = e E_0 / (m_i \nu_{in})$ defines $\nu_{in}$. To ensure compatibility between the simulation results and the theory, we choose $\nu_{in}$ in $\fptheoy(v_y)$ such that
\begin{equation}
\int_{-\infty}^{\infty} v_y \fptheoy(v_y) \, dv_y = \sum_{v_y} v_y F'_y(v_y) \Delta v.
\label{matchbulk}
\end{equation}
The expression on the left-hand side of Eq. \eqref{matchbulk} is the first velocity moment of $\fptheoy$ which gives the theoretical bulk velocity of the ions. The expression on the right-hand side of Eq. \eqref{matchbulk} gives the bulk velocity of the simulated ions. By matching these two quantities, we ensure that the theoretical ion distribution function is representative of the condition in the simulated background ions to first order.

We numerically calculated both sides of Eq. \eqref{matchbulk} for $\EEcase$, $\MEcase$, and $\HEcase$. For all of these cases, $\nu_{in} = \qty{1082}{s^{-1}}$ satisfies Eq. \eqref{matchbulk} to within $\pm \qty{2}{\mps}$. On the other hand, $\nu_{in} = \qty{1050}{s^{-1}}$ satisfies Eq. \eqref{matchbulk} only to within $\pm \qty{22}{\mps}$. Therefore, the simulated background ions exhibit an ion-neutral collision rate of \qty{1082}{s^{-1}} and not \qty{1050}{s^{-1}}.

Table \ref{tab:2} shows the matching bulk velocities for $\nu_{in} = \qty{1082}{s^{-1}}$, while Table \ref{tab:3} shows the bulk velocity mismatch for $\nu_{in} = \qty{1050}{s^{-1}}$. The choice of $\nu_{in}$ is irrelevant for $\NEcase$, since the theoretical ion distribution function is an undisturbed Maxwellian. 

\subsection{Comparison of the Theoretical and Simulated Ion Distribution Functions}

Figure \ref{fig:2a} compares the theoretical and simulated ion distribution functions in the Pedersen direction, that is, the direction parallel to $\vec{E_0}$. Equation \eqref{fptheo} gives the theoretical ion distribution functions in the Pedersen direcrtion. Equation \eqref{fp} gives the normalized ion distribution functions for the simulation results. Figure \ref{fig:2a} also includes the Maxwellian distribution functions which have the same bulk velocities as the simulation results but assume the neutral thermal velocity of \qty{287}{\mps}. 

In the Pedersen direction, both the theory and the simulation results show ion heating beyond Maxwellian, although the exact shapes of the distribution differ between the theory and the simulation results. The theoretical ion distribution functions are further right-skewed when compared to the simulation, although both are right-skewed when compared to the Maxwellian.

Figures \ref{fig:2b} and \ref{fig:2c} show the simulated ion distribution functions in directions perpendicular to $\vec{E_0}$. For comparison, the figure includes the undisturbed Maxwellian function which assumes the neutral temperature as the ion temperature. As mentioned in Section \ref{sec:yakov}, the theory assumes this undisturbed Maxwellian distribution in the perpendicular directions. The simulation results show ion heating beyond the neutral temperature, especially when $E_0$ is high. Figures \ref{fig:2b} and \ref{fig:2c} are largely identical due to symmetry.

Table \ref{tab:2} reports the bulk and the thermal velocities from the theory and the simulation. Section \ref{sec:disscussion} discusses the results in more details.
\def\fignum{2}
\setcounter{figure}{\fignum}
\setcounter{subfigure}{0}
\begin{subfigure}
\setcounter{figure}{\fignum}
\setcounter{subfigure}{0}
    \centering
    \begin{minipage}[b]{\sfwidth}
    	\caption{}
        \includegraphics[width=\linewidth]{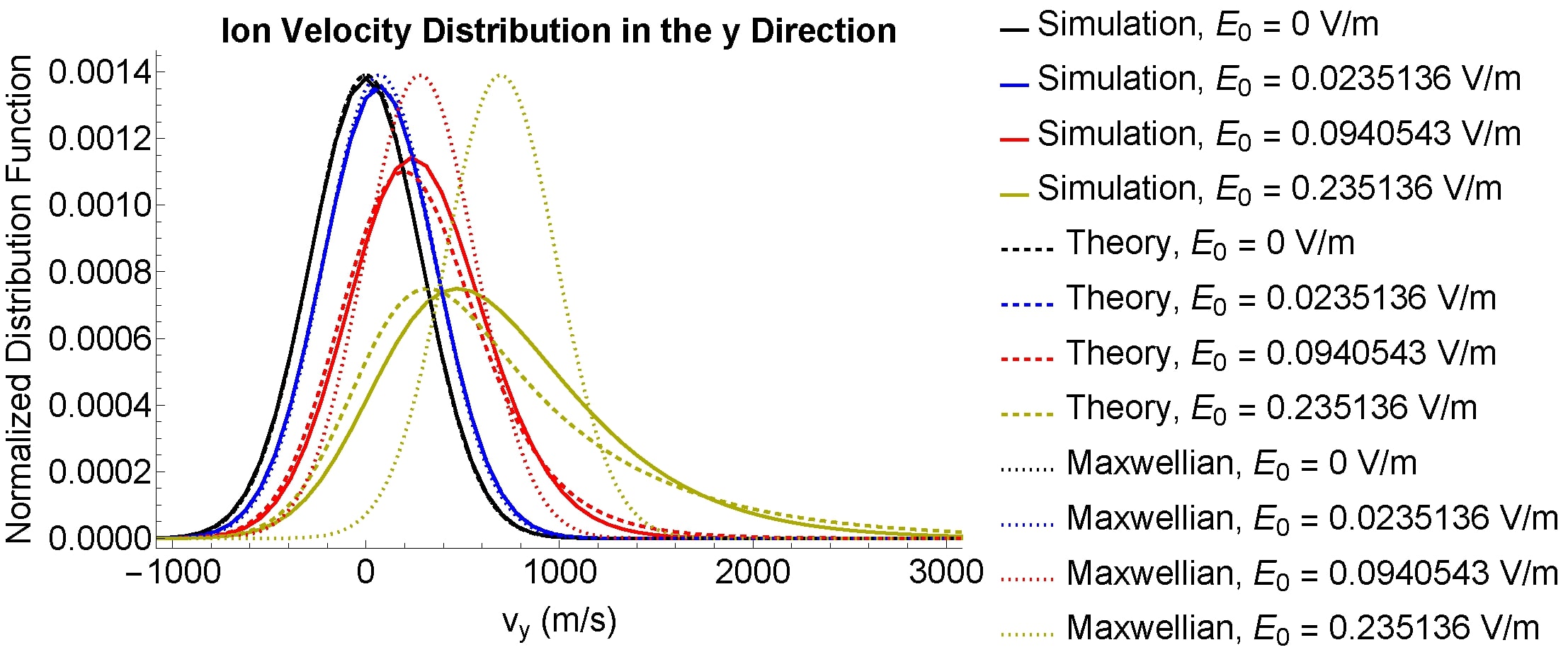}
        \label{fig:2a}
    \end{minipage}  
\setcounter{figure}{\fignum}
\setcounter{subfigure}{1}
    \begin{minipage}[b]{\sfwidth}
    	\caption{}
        \includegraphics[width=\linewidth]{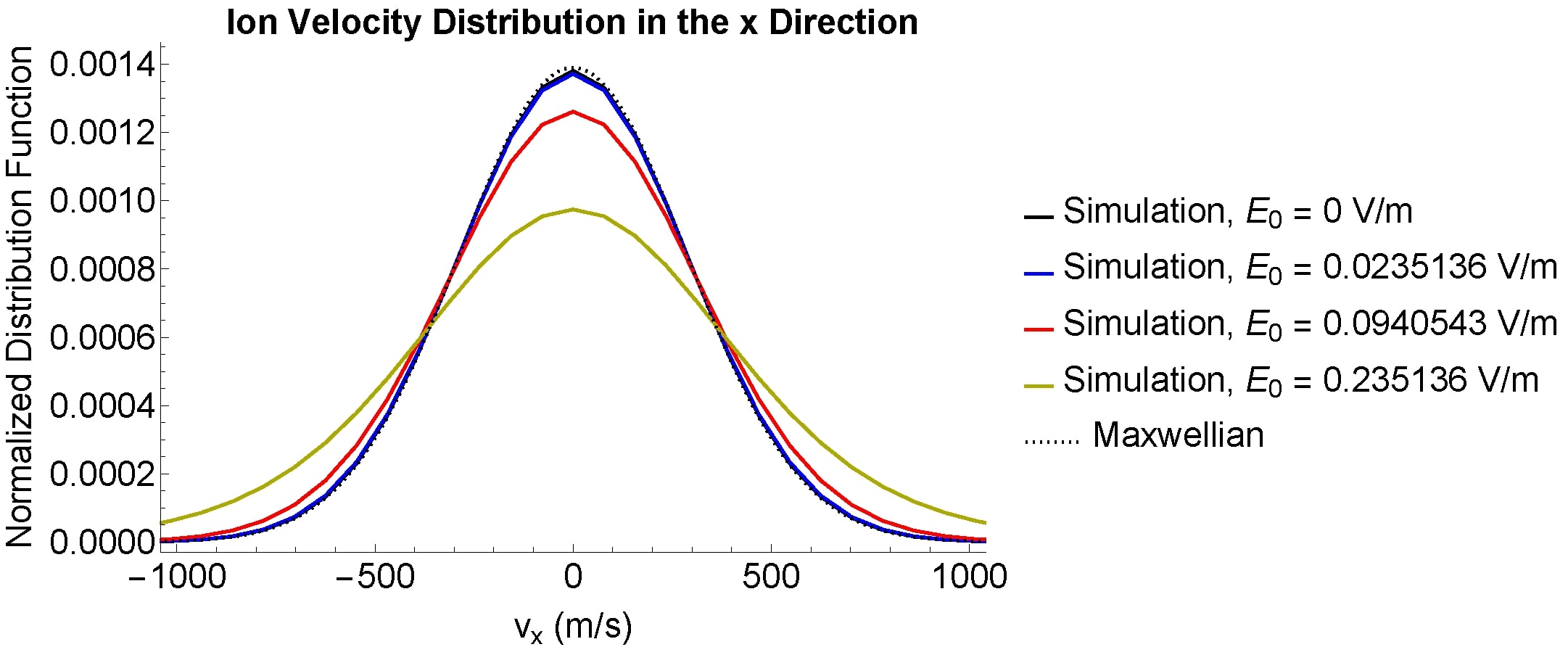}
        \label{fig:2b}
    \end{minipage}  
\setcounter{figure}{\fignum}
\setcounter{subfigure}{2}
    \begin{minipage}[b]{\sfwidth}
    	\caption{}
        \includegraphics[width=\linewidth]{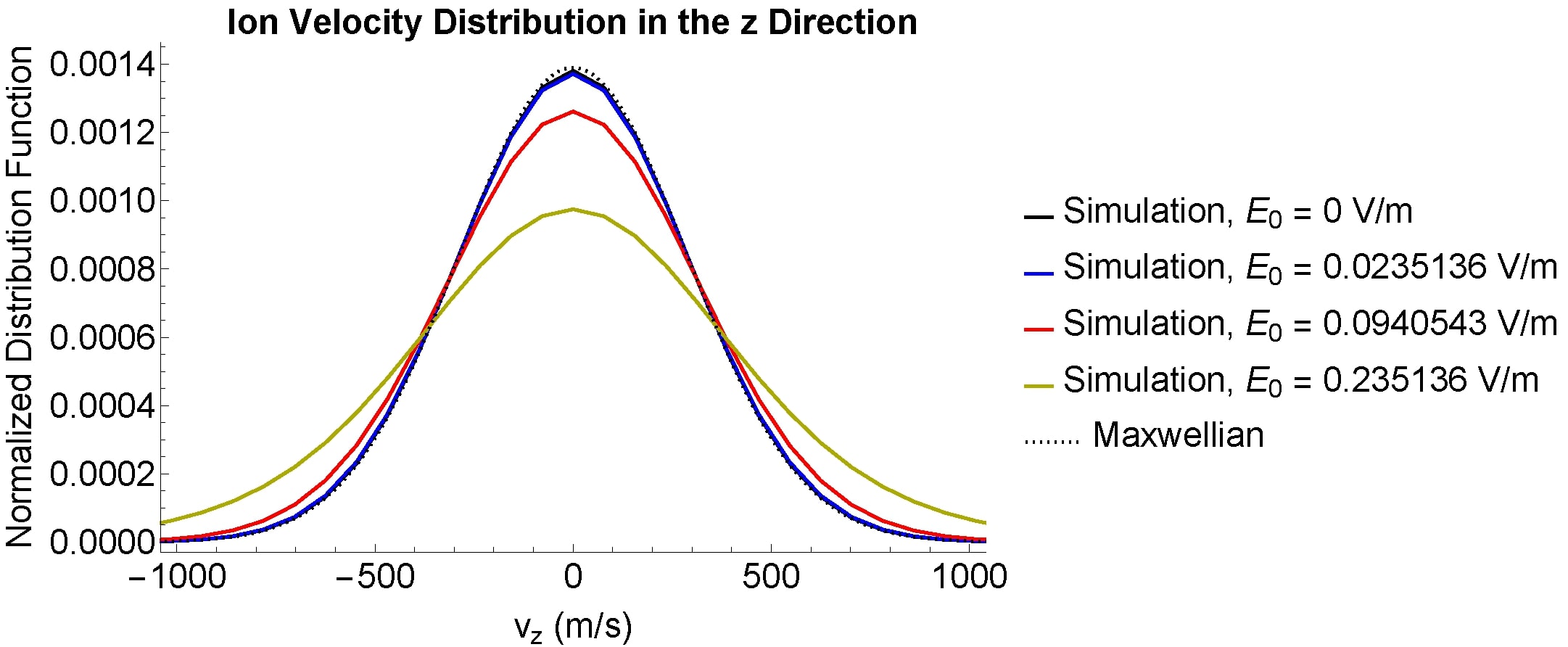}
        \label{fig:2c}
    \end{minipage}
\setcounter{figure}{\fignum}
\setcounter{subfigure}{-1}
    \caption{Comparison of the simulated (solid) and theoretical (dashed) ion velocity distribution functions. The Maxwellian functions (dotted) are included for comparison. The imposed electric field strengths are $\NEcase$ (black), $\EEcase$ (blue), $\MEcase$ (red), and $\HEcase$ (yellow). \textbf{(A)} Comparison of the ion velocity functions in the Pedersen direction. \textbf{(B, C)} Comparison of the ion velocity functions in the directions perpendicular to $\vec{E_0}$. The theory assumes an undisturbed Maxwellian in \textbf{(B)} and \textbf{(C)}. Due to symmetry, \textbf{(B)} and \textbf{(C)} are largely identical.}
    \label{fig:2}
\end{subfigure}
\begin{table}
\caption{Bulk velocities, directional thermal velocities, and total thermal energies for $\nu_{in} = \qty{1082}{\per\second}$. $\vybulk$ and $\vthy$ are the bulk velocity and the thermal velocity in the Pedersen direction respectively. $\vthj = \vthx = \vthy$ is the thermal velocity in the directions perpendicular to $\vec{E_0}$. $\sum v_{th}^2 = \vthy^2 + 2\vthj^2$ is the total thermal energy per ion mass. The last column shows the total energy ratio between the theory and the simulation results.}
\centering
\setlength{\extrarowheight}{3pt}
\begin{tabular}{|l|c|c|c|c|c|}
\hline 
Case 			& $\vybulk$				& $\vthy$				& $\vthj$				& $\sum v_{th}^2$													& Theory/Simulation \\
					&	(\mps)					&	(\mps)				& (\mps)				& (\unit[per-mode = symbol]{\joule\per\kilo\gram})		& Energy Ratio \\	
\hline
\multicolumn{6}{|l|}{\textit{$\NEcase$}} \\
\hline
Simulation	& 0		& 289	& 289	& 250563		& \multirow{2}{*}{0.9862}  \\ 
\cline{1-5}
Theory			& 0		& 287	& 287	& 247107		& \\ 
\hline
\multicolumn{6}{|l|}{\textit{$\EEcase$}} \\
\hline
Simulation 	& 70 	& 294 	& 291	& 255798		& \multirow{2}{*}{0.9849}   \\ 
\cline{1-5} 
Theory 			& 70 	& 295 	& 287 	& 251763		& \\ 
\hline
\multicolumn{6}{|l|}{\textit{$\MEcase$}} \\
\hline 
Simulation 	& 279 	& 358 	& 317 	& 329142		& \multirow{2}{*}{0.9864} \\ 
\cline{1-5} 
Theory 			& 279 	& 400 	& 287 	& 324738		& \\ 
\hline
\multicolumn{6}{|l|}{\textit{$\HEcase$}} \\
\hline
Simulation 	& 697 	& 606 	& 444 	& 761508		& \multirow{2}{*}{0.9611} \\ 
\cline{1-5} 
Theory 			& 696 	& 753 	& 287 	& 731747		&	\\ 
\hline 
\end{tabular}
\label{tab:2}
\end{table}

\begin{table}
\caption{Bulk velocities, directional thermal velocities, and total thermal energies for $\nu_{in} = \qty{1050}{\per\second}$. $\vybulk$ and $\vthy$ are the bulk velocity and the thermal velocity in the Pedersen direction respectively. $\vthj = \vthx = \vthy$ is the thermal velocity in the directions perpendicular to $\vec{E_0}$. $\sum v_{th}^2 = \vthy^2 + 2\vthj^2$ is the total thermal energy per ion mass. The last column shows the total energy ratio between the theory and the simulation results.}
\centering
\setlength{\extrarowheight}{3pt}
\begin{tabular}{|l|c|c|c|c|c|}
\hline 
Case 			& $\vybulk$				& $\vthy$				& $\vthj$				& $\sum v_{th}^2$													& Theory/Simulation \\
					&	(\mps)					&	(\mps)				& (\mps)				& (\unit[per-mode = symbol]{\joule\per\kilo\gram})		& Energy Ratio \\	
\hline
\multicolumn{6}{|l|}{\textit{$\NEcase$}} \\
\hline
Simulation	& 0		& 289	& 289	& 250563		& \multirow{2}{*}{0.9862}  \\ 
\cline{1-5}
Theory			& 0		& 287	& 287	& 247107		& \\ 
\hline
\multicolumn{6}{|l|}{\textit{$\EEcase$}} \\
\hline
Simulation 	& 70 	& 294 	& 291	& 255798		& \multirow{2}{*}{0.9861}   \\ 
\cline{1-5} 
Theory 			& 72 	& 296 	& 287 	& 252244		& \\ 
\hline
\multicolumn{6}{|l|}{\textit{$\MEcase$}} \\
\hline 
Simulation 	& 279 	& 358 	& 317 	& 329142		& \multirow{2}{*}{1.0010} \\ 
\cline{1-5} 
Theory 			& 287 	& 406 	& 287 	& 329476		& \\ 
\hline
\multicolumn{6}{|l|}{\textit{$\HEcase$}} \\
\hline
Simulation 	& 697 	& 606 	& 444 	& 761508		& \multirow{2}{*}{1.0005} \\ 
\cline{1-5} 
Theory 			& 718 	& 773 	& 287 	& 761914		&	\\ 
\hline
\end{tabular}
\label{tab:3}
\end{table}

\section{Discussion}
\label{sec:disscussion}

In this section, we mostly discuss discrepancies between the analytical results of Section \ref{sec:yakov} and the PIC simulations. On the one hand, the analytical model (hereinafter referred to as 'theory') is not perfectly accurate because it is based on the oversimplified BGK collision model. As a result, the theoretical 3-D shape of the ion distribution function turns out to be less accurate than the PIC one (overdistorted in the electric field direction and undisturbed Maxwellian in the two perpendicular directions). On the other hand, the integrated fluid characteristics, such as the ion bulk velocity and the total ion temperature, elevated due to frictional heating by the external electric field, should be accurately represented by this theory, even in the cases of very strong electric fields that result in efficient distortions of the ion distribution function. If there still remain small discrepancies, then this may be attributed to not perfectly matching collision rates and to the fact that the velocity integration of the PIC determined ion distribution function is performed within an artificially restricted velocity domain. This is especially relevant to the strongly distorted ion distribution function when its high-energy tail can include a noticeable fraction of particles.

\subsection{Thermal Velocity Mismatch Between the Theory and the Simulation Results}
\label{sec:vthresults}

The simulated ion distribution functions show different thermal profiles from those predicted by the theory. 

\subsubsection{Definition of Thermal Velocity}

For the theory, the thermal velocity in the Pedersen direction is defined in terms of the second velocity moment of the ion distribution function:
\begin{equation}
\vthy^{\mathrm{Theory}} = \sqrt{\int_{-\infty}^{\infty}\left(v_y - \left\langle v_y \right\rangle\right)^2 \fptheo_y (v_y)\, dv_y},
\label{vthdeftheo}
\end{equation}
where $\fptheoy$ is the normalized ion distribution function from Eqn. \eqref{fptheo}, and $\left\langle v_y \right\rangle$ is the ion bulk velocity in the Pedersen direction as given in Table \ref{tab:2}. In directions perpendicular to $\vec{E_0}$, the thermal velocity is equal to the neutral thermal velocity $v_T$, since the theory does not account for heating in these directions and assumes an undisturbed Maxwellian. 

For the simulation results, the thermal velocity in direction $i$ is given by
\begin{equation}
v_{th, i} = \sqrt{\sum_{v_i}\left(v_i - \left\langle v_i \right\rangle\right)^2 F'_i (v_i)\, dv_i},
\end{equation}
where $i$ is either $x$, $y$, or $z$; $F'_i$ is the normalized ion distribution function from Eqn. \eqref{fp}; and $\left\langle v_i \right\rangle$ is the ion bulk velocity in direction $i$ as given in Table \ref{tab:2}.

Table \ref{tab:2} shows the mismatch in directional heating between the theory and the simulation results. Section \ref{sec:dissvthperp} discusses ion heating in directions perpendicular to $\vec{E_0}$, while Section \ref{sec:dissvthped} discusses ion heating in the Pedersen direction.

\subsubsection{Underestimation of the Thermal Velocity in the Directions $\perp \vec{E_0}$}
\label{sec:dissvthperp}

The theory underestimates the ion heating in the directions perpendicular to $\vec{E_0}$. In the $x$ and $z$ directions, the theory predicts the ion thermal velocity of \qty{287}{\mps} which is equal to the neutral thermal velocity $v_T$. 

The simulation for $\NEcase$ gives the thermal velocity of \qty{289}{\mps}, which is \qty{2}{\mps} higher than the predicted value. Although a 0.7 percent difference is not large, an explanation exists for the discrepancy. Since we initialized the ion thermal velocity at \qty{575}{\mps}, they needed time cool down to the neutral thermal velocity through collisions with the neutrals. The simulation time might not have been enough for the ions to completely reach steady state.

For larger values of $E_0$, the simulation shows an increase in the ion thermal velocity, whereas the theoretical thermal velocity remains at \qty{287}{\mps}. The theory assumes an undisturbed Maxwellian in the directions perpendicular to $\vec{E_0}$, so it does not account for ion heating in these directions. The simulation shows that ion heating is more intense for larger values of $E_0$. In the most intense case $\HEcase$, the simulated thermal velocity reaches as high as \qty{444}{\mps} or about 1.5 times the undisturbed value.

\subsubsection{Overestimation of the Thermal Velocity in the Direction $\parallel \vec{E_0}$}
\label{sec:dissvthped}

The theory overestimates the heating in the Pedersen direction. In the $y$ direction, the theory predicts higher ion thermal velocities for higher values of $E_0$. Table \ref{tab:2} shows the theoretical predictions of the thermal velocities as well as the simulation results.

The simulation for $\NEcase$ gives the thermal velocity of \qty{289}{\mps}, which is \qty{2}{\mps} higher than the predicted value. Section \ref{sec:dissvthperp} discusses the discrepancy.

For larger values of $E_0$, both the theory and the simulation show increased ion thermal velocities beyond the neutral thermal velocity. However, the theory and the simulation results disagree on the exact amount of the heating. The simulation shows that ion heating is less intense in the Pedersen direction than theory suggests. The discrepancy is larger for larger values of $E_0$. In the most intense case $\HEcase$, the simulated thermal velocity only reaches \qty{606}{\mps} or just 80 percent of the theoretical value, \qty{753}{\mps}. 

\subsubsection{Angular Scattering of Ions Due to Elastic Collisions with the Neutrals}

The major difference between the theory and the simulation is the angular scattering of ions in 3-D. The theory models ion heating only in the Pedersen direction; it does not account for ions scattering into the directions perpendicular to $\vec{E_0}$. On the other hand, the PIC code is able to capture the physics of ion scattering in 3-D. Angular scattering causes ion heating to be more isotropic in the simulation. The theory underestimates the heating in the directions it does not account for, while at the same time overestimating in the direction it does account for. 

We expect the total ion thermal energy to be the same between the theory and the simulation.  Section \ref{sec:dissenergy} compares the total energy between the theory and the simulation.

\subsection{Discrepancy in Total Energy Between the Theory and the Simulation Results}
\label{sec:dissenergy}

The total ion thermal energy differs between the theory and the simulation results. Table \ref{tab:2} gives the total thermal energy per ion mass as well as the total thermal energy ratio between the theory and the simulation results. While the ratios are close to one, the total thermal energy from the theory is consistently lower than the total thermal energy from the simulation. Larger values of $E_0$ exhibit larger energy discrepancies than smaller values of $E_0$. In the most intense case $\HEcase$, the theory captures 96.11 percent of the total simulated energy, while in the less intense case $\EEcase$, the theory captures as much as 98.85 percent of the total simulated energy.

A possible explanation for the discrepancy in total energy is our choice of $\nu_{in}$ as described in Section \ref{sec:nu}. The theoretical IDF depends on $\nu_{in}$ in the Pedersen direction. We chose $\nu_{in}$ retroactively such that the theory matches the simulation results to first order. Table \ref{tab:3} shows a hypothetical situation in which the theory uses $\nu_{in} = \qty{1050}{s^{-1}}$ as its ion-neutral collision rate instead of the analytical value of \qty{1082}{s^{-1}}. As seen by the mismatch in the bulk velocity, $\nu_{in} = \qty{1050}{s^{-1}}$ does not satisfy Eq. \eqref{matchbulk}. However, $\nu_{in} = \qty{1050}{s^{-1}}$ shows greater agreement with the simulation results in terms of the total thermal energy. This is especially true for larger values of $E_0$ where the energy in the theory and the simulation is almost identical.

Comparing Table \ref{tab:2} and Table \ref{tab:3} shows how sensitive the theoretical IDF is to the value of $\nu_{in}$. We expect the theory to preserve the total thermal energy of the background ions while also giving the correct ion bulk velocity. The theory is able to do both within a margin of error.

\subsection{Distortion of the Ion Distribution Function in the Equatorial E-Region}

A typical DC electric field strength in the equatorial E-region is $\EEcase$. Figure \ref{fig:2} shows only a small distortion in the ion distribution function for $\EEcase$. Table \ref{tab:2} gives the bulk and thermal velocities for $\EEcase$. 

In the Pedersen direction, the theory predicts a thermal velocity of \qty{295}{\mps}, while the simulation shows a thermal velocity of \qty{294}{\mps}. In the directions perpendicular to the Pedersen direction, the simulation shows a thermal velocity of \qty{291}{\mps}. These numbers are not so different from the Maxwellian thermal velocity which is \qty{287}{\mps}.

The background ion distribution in the equatorial E-region is not likely to distort much from Maxwellian, because the electric field is not strong enough. Both the theory and the simulation show that the distortion is stronger when $E_0$ is higher. In the Earth's ionosphere, the distortion will be stronger in the auroral E-region where the DC electric field is more intense than the equatorial E-region, especially during periods of geomagnetic storms.

\section{Conclusions}
\label{sec:conclusions}

We developed a collisional plasma kinetic model for the E-region background ions using the simple BGK collision operator, see Section ~\ref{sec:yakov}. This simplified analytic model result in the ion distribution function (IDF) which is distorted in the direction of the external DC electric field $\vec{E_0}$ (the Pedersen direction), while in the two perpendicular directions the velocity distribution remains undisturbed Maxwellian, as described by Eqs.~(\ref{F_0})--(\ref{F_0(V_x)}). The reason for this extreme anisotropy lies in the fact that the BGK collisional operator does not include any ion angular scattering in the velocity space. At the same time, even this simplified model provides accurate values for the total Pedersen drift velocity and, given equal masses of the colliding ions and neutrals, for the total effective ion temperature elevated by the frictional heating. Under a sufficiently intense external electric field, the IDF is skewed in the direction of $\vec{E_0}$, so that a strong tail of superthermal-energy ions forms.

We compared this simplified model to the PIC simulation, see Section ~\ref{sec:PIC}. The simulation shows less ion heating in the Pedersen direction and more ion heating in the perpendicular directions when compared to the analytic model. The difference in the thermal distribution is due to the ion angular scattering which, unlike the model, is present in the PIC code. However, there is also a small difference in the total thermal energy between the model and the simulation. This difference can be attributed to the choice of the ion-neutral collision rate. We have shown that the model is sensitive to the choice of the ion-neutral collision rate. The more accurate IDF determined by the PIC simulation is somewhere between the analytically determined IDF and the Pedersen-shifted Maxwellian distribution whose temperature equals the total elevated ion temperature. The latter, however, does not show any IDF skewness which is present in both analytical model and PIC simulations.

For the typical electric field strength of the equatorial E-region, the background ion distribution function is well-represented by the shifted and heated Maxwellian function. The situation may be very different at high latitudes where a strong external field may be present during periods of geomagnetic storms. Both the model and the PIC simulation show that in these cases, the background ion velocity distribution can distort significantly from any Maxwellian.

\section*{Conflict of Interest Statement}

The authors declare that the research was conducted in the absence of any commercial or financial relationships that could be construed as a potential conflict of interest.

\section*{Funding}
This work was funded by NASA Grants 80NSSC21K1322 and 80NSSC19K0080.

\section*{Acknowledgments}
Computational resources were provided by the Anvil supercomputer through the NSF/ACCESS program.

\bibliographystyle{Frontiers-Harvard} 
\bibliography{nonmaxwellian_electrojet_ions}



\end{document}